\documentclass[a4paper,11pt]{article}
\pdfoutput=1 %

\usepackage{jinstpub} %
\PassOptionsToPackage{hyphens}{url}
\appto\UrlBreaks{\do\/}
\appto\UrlBigBreaks{\do\/}

\usepackage{lineno}
\title{\boldmath GBTX emulator for development and special versions of GBT-based readout chains}

\author[a]{W.M. Zabołotny,\note{Corresponding author.}}
\author[a]{A.P. Byszuk,}
\author[b]{D. Dementev,}
\author[c]{D. Emschermann,}
\author[a]{M. Gumiński,}
\author[a]{M. Kruszewski,}
\author[a]{P. Miedzik,}
\author[a]{K. Poźniak,}
\author[a]{R. Romaniuk,}
\author[c]{C.J. Schmidt}
\author[b]{and M. Shitenkov}

\affiliation[a]{Institute of Electronic Systems, Warsaw University of Technology,\\
             Nowowiejska 15/19, 00-665 Warszawa, Poland}
\affiliation[b]{Veksler and Baldin Laboratory of High Energy Physics, Joint Institute for Nuclear Research,\\
             Baldin str. 6, Dubna 141980, Russian Federation}
\affiliation[c]{GSI - Helmholtzzentrum für Schwerionenforschung GmbH,\\
             Planckstraße 1, 64291 Darmstadt, Germany}

\emailAdd{wzab@ise.pw.edu.pl}

\abstract{The GBTX ASIC is a standard solution for providing fast control and data readout for radiation detectors used in HEP experiments. However, it is subject to export control restrictions due to the usage of radiation-hard technology. An FPGA-based GBTX emulator (GBTxEMU) has been developed to enable the development of GBT-based readout chains in countries where the original GBTX cannot be imported. Thanks to utilizing a slightly modified GBT-FGPA core, it maintains basic compatibility with standard GBT-based systems. The GBTxEMU also may be an interesting solution for developing GBT-based readout chains for less demanding experiments.
}


\arxivnumber{2109.11591} %

\proceeding{22$^{\text{nd}}$ International Workshop on Radiation Imaging Detectors\\
  June 27 – July 1, 2021\\
  Ghent, Belgium}

\usepackage[frozencache=true,cachedir=minted-cache]{minted}
\usepackage{multicol}
\usepackage{cprotect}

\newcommand{\wzurl}[1]{\href{#1}{\url{#1}}}
\newcommand{\wzcode}[1]{{\tt \bf #1}}

\newcommand{\wzlistingforth}[1]{\inputminted[fontfamily=tt, style=bw]{forth}{#1}}

\begin{document}
\maketitle
\flushbottom

\section{Introduction}

The GBTX ASIC~\cite{url-gbtx} is a standard solution for fast control and data readout for radiation detectors used in HEP experiments \cite{mitra_trigger_2019,marin_gbt-fpga_2015,agwb-spie,dementev_fast_2021}.
Together with the dedicated GBT-FPGA~\cite{marin_gbt-fpga_2015} core suitable for implementation in FPGA,
it can be used for time-deterministic transmission of commands in the downlink direction
and transmission of command responses and readout data in the uplink direction.
For communication, a dedicated high-speed 4.8~Gb/s GBT Link protocol~\cite{Moreira2007332} is used.
Connection to the Front-End ASICs is made via convenient SPI-Like E-Link interfaces.

Many readout systems have been created based on the GBTX ~\cite{antonioli_design_2016, naik_-detector_2017, fernandez_prieto_phase_2020, lehnert_gbt_2017}.
For the STS and MUCH detectors in the CBM experiment, dedicated front-end ASICs~\cite{kasinski_sts-xyter_2014} have been developed, together with
a special HCTSP protocol~\cite{kasinski_protocol_2016} oriented on communication via AC-coupled links.

Unfortunately, the GBTX is not widely available. It is not a "custom of the shelf" (COTS) part. Additionally, due to the usage
 of the radiation-hard technology, it is subjected to export restrictions and unavailable in certain countries.

Therefore, in many international experiments, certain collaboration members cannot develop
GBTX-based readouts in their countries.
The usage of GBTX chips in all development sites also may not be justified from an economic point of view.

The GBTX emulator (GBTxEMU) board has been developed to eliminate the mentioned limitations.

\section{Concept of the GBTxEMU board}

The concept of the GBTxEMU board was based on a few assumptions:
\begin{itemize}
	\item It should provide essential functionalities of the GBTX ASIC,
	\item It should be widely available (no export restrictions on components),
	\item It should be easily affordable (possibly based on cheap, mass-produced COTS components),
\end{itemize}

A relatively cheap commercial board
from Trenz~\cite{url-trenz-el} TE0712~\cite{url-te0712}
with an Artix 7 FPGA was selected to fulfill the last two requirements.
Of course, the GBTxEMU should emulate not the GBTX alone but the whole GBTX-based readout board.
Therefore, the TE0712 board had to be supplemented with an additional baseboard
equipped with necessary interfaces and other infrastructure.
For example, obtaining the reference clock from the clock recovered from the GBT link
requires a jitter cleaner.

\begin{figure}[htbp]
	\centering %
	\includegraphics[width=.9\textwidth]{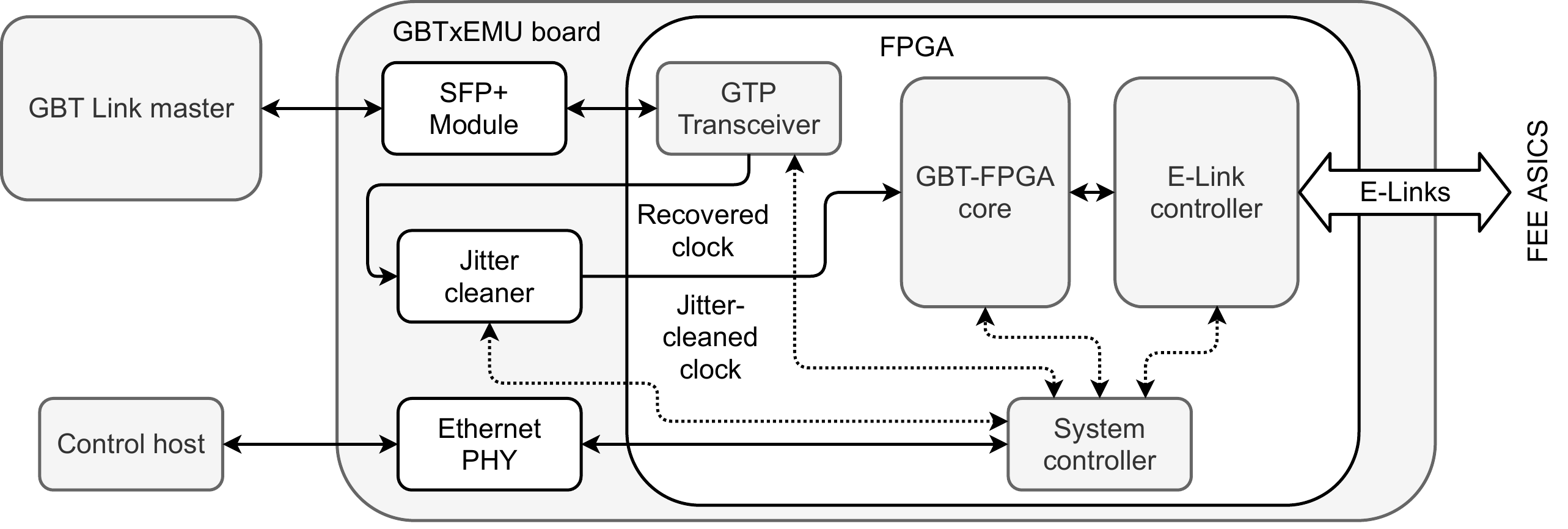}
	\caption{\label{fig:gen-concept} The general concept of the GBTxEMU board.}
\end{figure}

The general concept of the GBTxEMU board is shown in Figure~\ref{fig:gen-concept}.

\section{Hardware platforms for the GBTxEMU board}

\begin{figure}[htbp]
	\centering %
	\includegraphics[width=.7\textwidth]{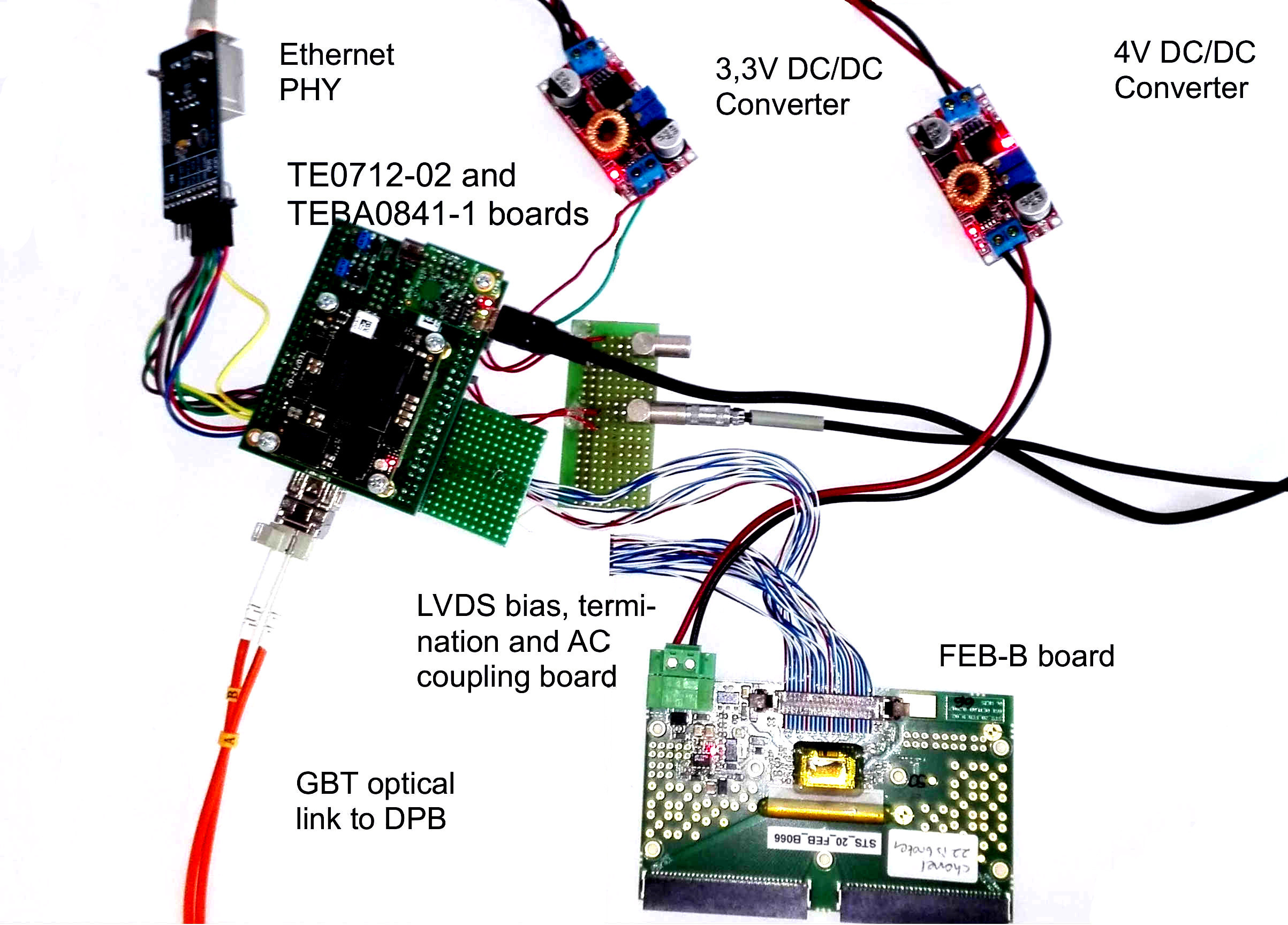}
	\caption{\label{fig:prototype} The first prototype of GBTxEMU, based on TE0712 and TEBA0841 boards. Picture from the presentation~\cite{zabolotny_gbtxemu_2019}.}
\end{figure}

The first prototype of the GBTxEMU~\cite{cbm-prog-rep-2018-WUT-gbtxemu}
 (see Figure~\ref{fig:prototype}) used the TEBA0841-1 baseboard by Trenz~\cite{url-teba841},
providing an SFP+ cage for the GBT Link transceiver and gold-pin header for E-link connections.
 Additionally,
a 100~Mb/s DP83848 Ethernet PHY module~\cite{url-dp83848-module} was connected.

The Si5338 clock generator, controlled by the FPGA, was used as a jitter cleaner in that version.

\begin{figure}[htbp]
	\centering %
	\includegraphics[height=4 cm]{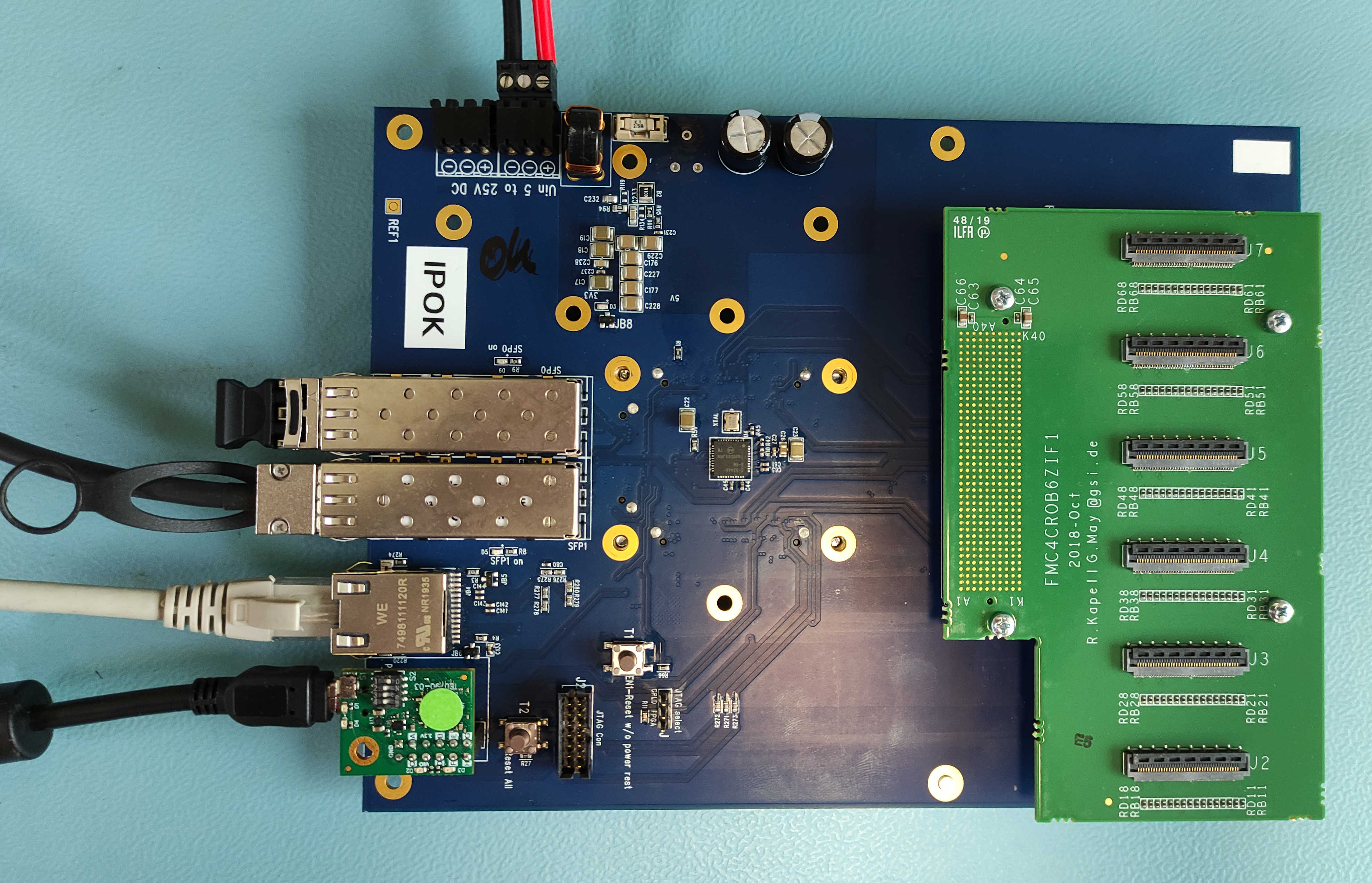}
	\qquad
	\includegraphics[height=4 cm]{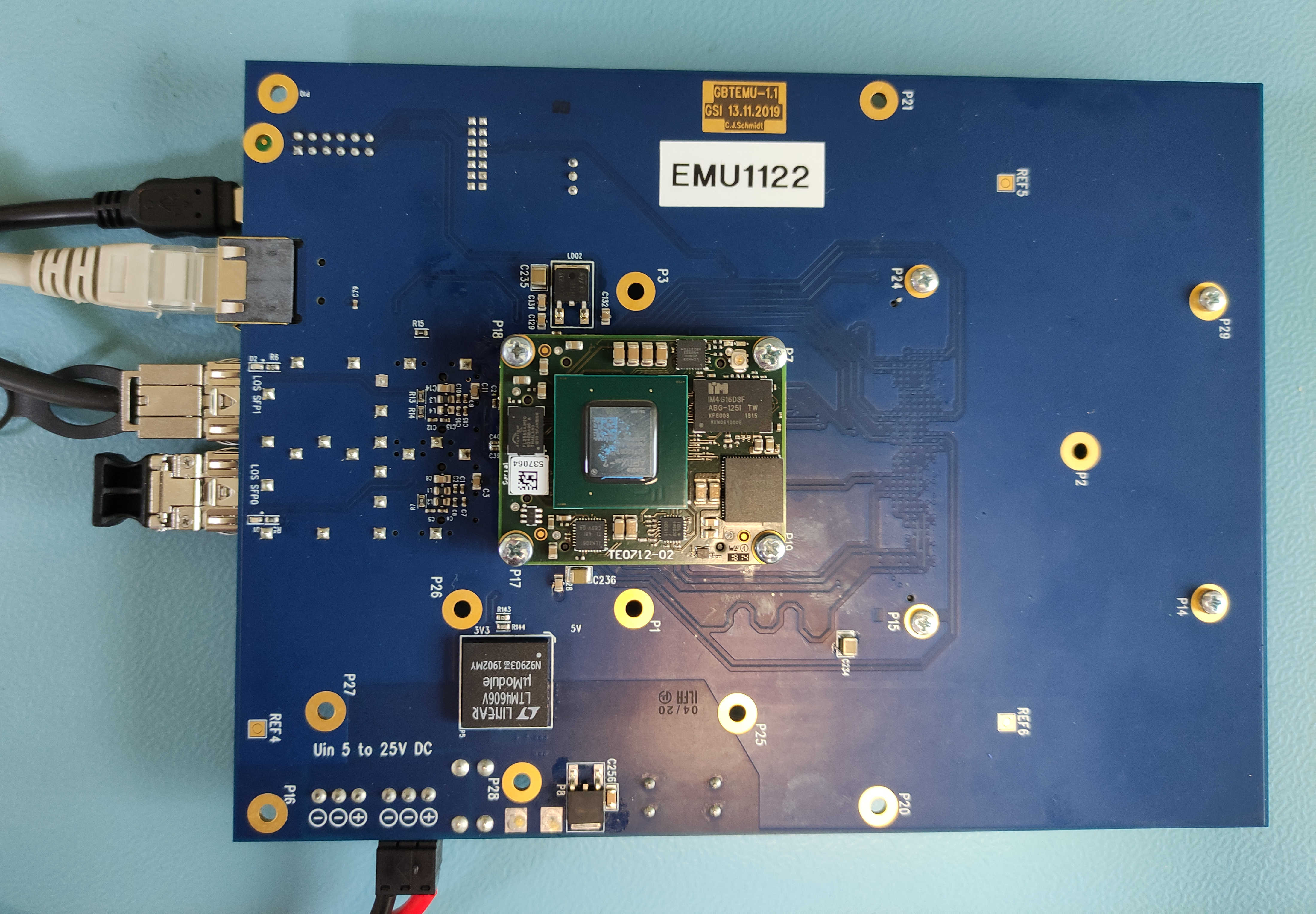}
	\caption{\label{fig:emu-v1} The first version of the GBTxEMU board developed at GSI. 
		On the left -- the top side with the FMC equipped with ZIF connectors for FEBs.
		On the right -- the bottom side with a TE0712-2 module.
	}
\end{figure}

Experiences gathered with the first prototype were used to develop the first version of the GBTxEMU board (shown in Figure~\ref{fig:emu-v1}) in GSI.
It is equipped with two SFP+ cages and the hardware jitter cleaner based on a Silabs Si5344 chip~\cite{url-si5344}.

That version was equipped with the FMC connector enabling the use of FMC boards
with various FEB connectors. Due to versatility, that version is a good solution for development and testing.

\begin{figure}[htbp]
	\centering %
	\includegraphics[width=.6\textwidth]{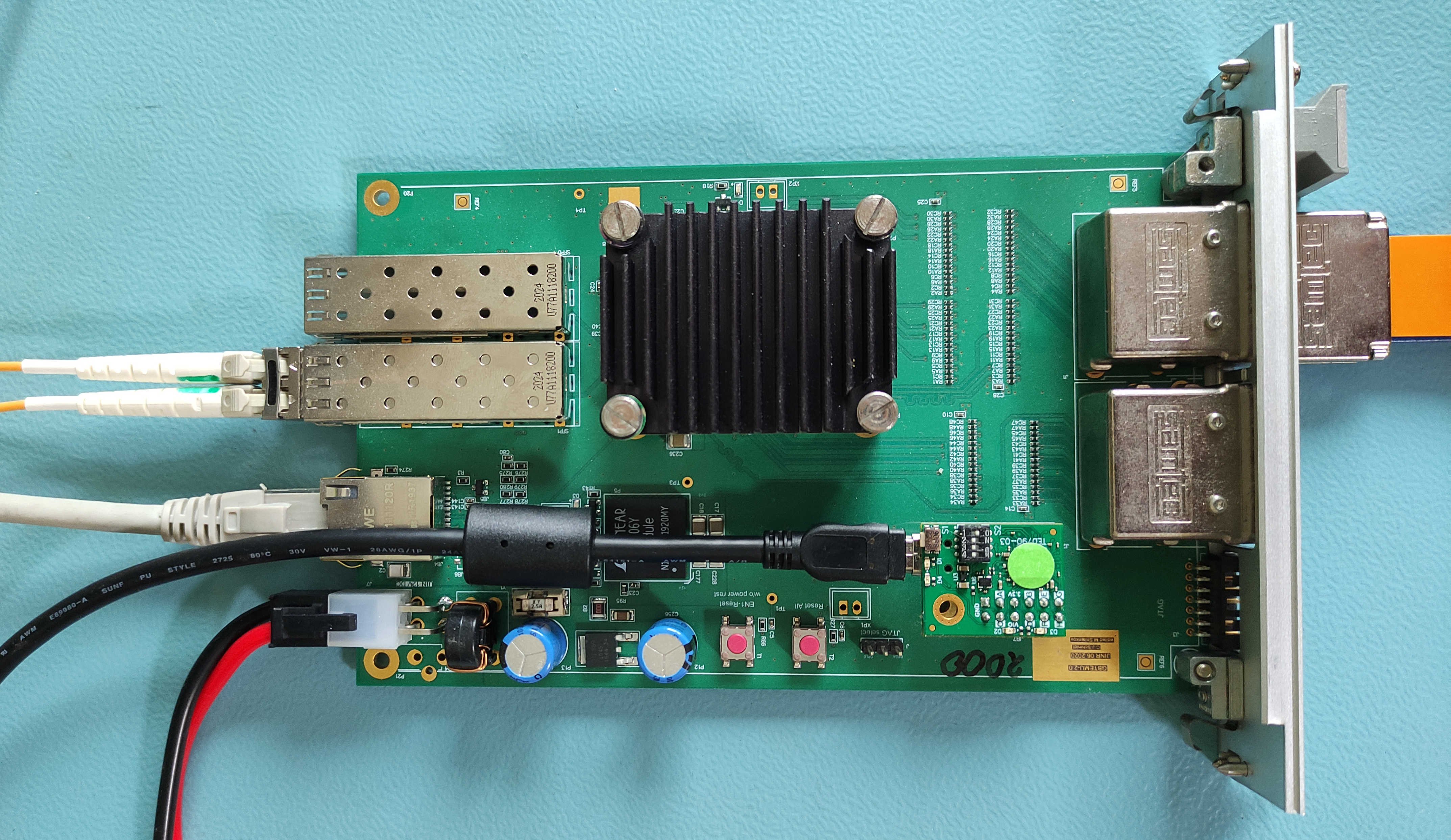}
	\caption{\label{fig:emu-v2} The second version of the GBTxEMU developed at JINR in the Eurocard format and with high-density Samtec connectors for FEBs. The TE0712-2 module is covered with a heat sink.}
\end{figure}

The GBTxEMU is also planned for use as a component of  the GBT-based
readout chain in the BM@N experiment~\cite{shitenkov_front-end_2021,dementev_fast_2021}.
For that purpose, the second version of the GBTxEMU board (see Figure~\ref{fig:emu-v2}) was prepared.
This version is designed in a Eurocard format.
The E-Links are connected to the high-density Samtec connectors instead of the FMC connector for better space utilization.

\section{Implementation of the GBTxEMU firmware}

\begin{figure}[htbp]
	\centering %
	\includegraphics[width=.7\textwidth]{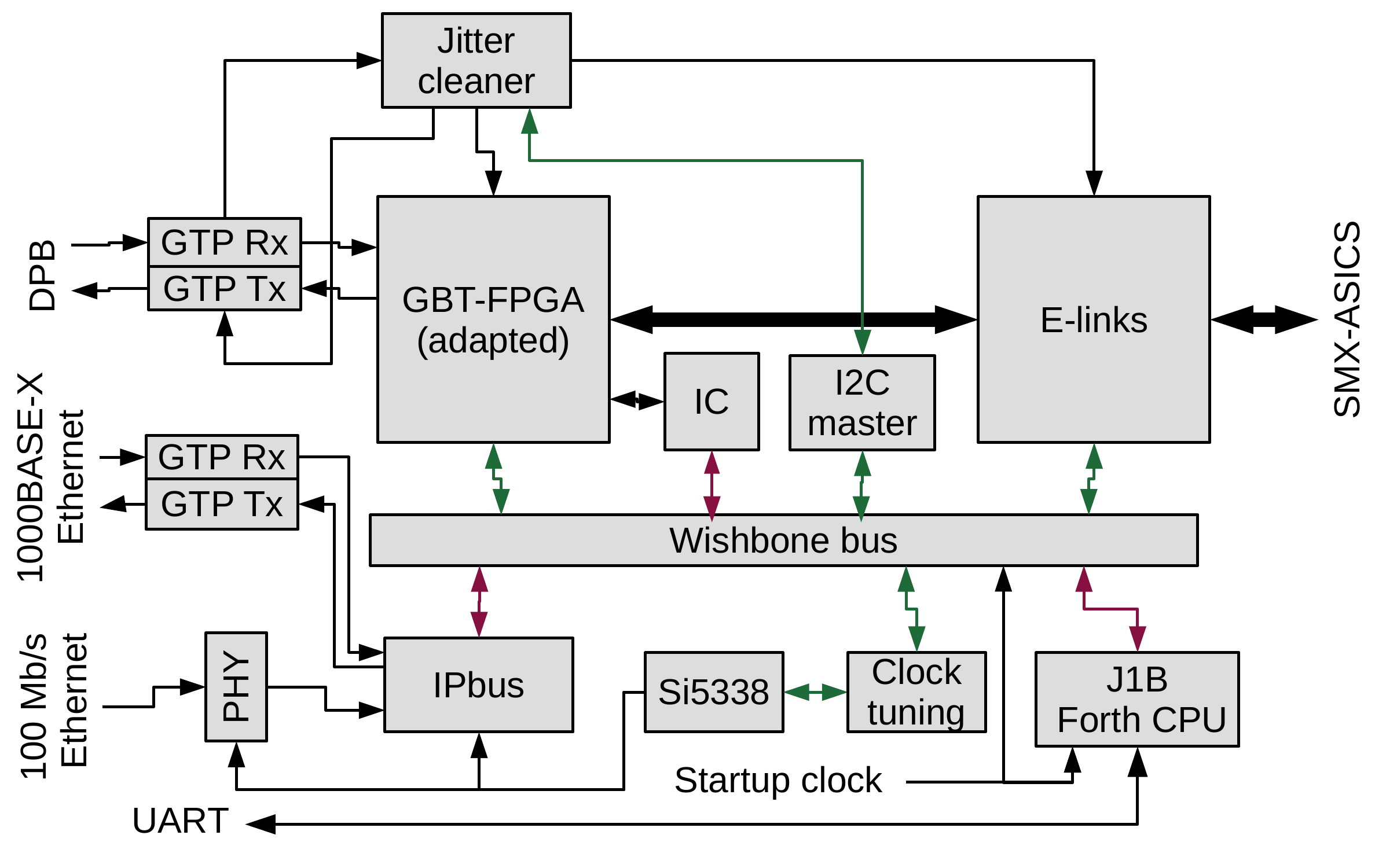}
	\caption{\label{fig:fw_block_diagram} Block diagram of the GBTxEMU firmware (based on~\cite{cbm-prog-rep-2019-WUT}).}
\end{figure}

The block diagram of the GBTxEMU firmware is shown in Figure~\ref{fig:fw_block_diagram}

 Because the main system clock is obtained by the jitter-cleaning of the clock recovered from the GBT Link
it means that certain initialization must be done before the main clock is available.
For that purpose the GBTxEMU is equipped with the J1B system controller using the additional ``boot clock''\footnote{The ``boot clock'' has a frequency of 10~MHz in the first batches of GBTxEMU and then it was replaced with
 20~MHz}.

\subsection{J1B system controller}

Initialization of the GBTxEMU requires configuring the clock generator, the jitter cleaner, and the GTP transceivers.
For that purpose, a simple synthesizable  J1B CPU~\cite{url-j1b} was implemented.
It is programmed in Forth~\cite{url-thinking-forth} language. Due to good code density, despite relatively small
32~kB RAM, it may perform quite complex initialization routines.
An example code that reads the board's MAC address from a UNI/O FLASH memory~\cite{url-unio-flash} is shown in Listing~\ref{lst:forth}.
\begin{listing}[htbp]
	\cprotect\caption{Example Forth code for the J1B, used to read the MAC address from the UNI/O FLASH memory.}\label{lst:forth}
	\rule{\columnwidth}{1pt}
	\begin{multicols}{2}
		\raggedright {\scriptsize \wzlistingforth{code/scio.fs}}
	\end{multicols}
\end{listing}

The Forth language is considered to be difficult to learn but offers efficient possibilities of interactive work.
It needs only a UART connection\footnote{It is also possible to emulate UART via JTAG. The J1B connected via JTAG-emulated UART is used in the KCU116\_J1B project~\cite{url-kcu116-j1b}.} to the operator's console. In the interactive mode, it is possible to create new procedures (called {\em words} in Forth)
and execute them. Thence Forth may be a perfect debugging tool. The captured word definitions may also be compiled and put into the GBTxEMU configuration
bitstream. The user-defined word \wzcode{cold} is executed after the system starts.

\subsection{GBTxEMU internal bus}
\label{sec:intbus}
Most functional blocks of the GBTxEMU are connected to the internal Wishbone bus. 
The bus is managed by the Address Generator for Wishbone (AGWB) system~\cite{agwb-spie} which assigns the addresses
for the internal registers,
 generates the VHDL code providing access to those registers,
 and generates the software modules for Forth and Python, enabling convenient access to those registers via their name.
 The address map may be automatically adjusted after modification of the register file.

The internal Wishbone bus has three masters. The first of them is J1B, described in the previous section and used for initialization and debugging.
The second is the IPbus~\cite{larrea_ipbus:_2015} master, which may control the GBTxEMU via the independent Ethernet interface (with 100~Mb/s or 1~Gb/s speed).
That requires that the board is initialized to the stage where the Ethernet connection is working.
Otherwise, it enables control and debugging using Python running on the computer, which is more convenient than Forth. 
The third master is the controller based on the GBT-SC core~\cite{url-gbt-sc}.
This master provides the best emulation of GBTX because control commands and responses
 are transmitted via the GBT link, just like in the case of the real GBTX.

\subsection{The GBT-FPGA core}
The GBT-FPGA core was developed by CERN~\cite{url-gbt-fpga}. It implements the essential GBTX functionalities in an FPGA.
The original implementation, however, does not support the Artix 7 FPGAs.
Therefore, certain modifications were necessary.

Adaptation of GBT-FPGA core was made based on a version prepared for Kintex 7
as it was the most similar.
The Artix 7 FPGAs use GTP transceivers~\cite{xlx-gtp-ug}, while Kintex 7 uses the GTX transceivers~\cite{xlx-gtx/h-ug}.
The necessary modifications were related to different widths of the internal
datapath,  internal clock frequencies, and configurations of the PLL blocks generating the transceiver clock.
The operation of GTP transceivers at 4.8~Gb/s (required for a standard GBT link) was achieved successfully.
 
The Artix 7 FPGAs offer lower performance than Kintex 7. Despite this, it was possible
to implement the forward error correction (FEC) in the downlink direction, which is essential for
reliable transmission of control commands to front-end ASICs. Also, the optimized latency mode
was possible in the downlink direction – it is necessary to synchronize the front-end ASICs.
In the uplink direction, the ported GBT-FPGA provides standard latency transmission in the
Widebus mode (without FEC).

\subsection{Implementation of E-Links}
The E-Link block is responsible for retrieving the appropriate bits from the downlink GBT frame and sending them serially via E-Links,
and simultaneously for receiving the bits from the E-Links and storing them into the uplink GBT frame.
The uplink direction is optimized for transmission of the hit data from the FEE ASIC. Thence, it is assumed that the uplink data are transmitted with a dual data rate (DDR).
Because the downlink in the FEC mode used by the GBTxEMU delivers one 80-bit downlink frame every 25~ns,
and the uplink in the Widebus mode accepts one 112-bit frame every 25~ns, 
the possible number of E-Links depends on the E-Link clock frequency $f_{\textrm{EL CLK}}$. 
It is limited by the number of bits available in the uplink frame and may be calculated according to the formula:
$
  N_{\textrm{E-Links}}=\textrm{ceil} \left( \frac{112 \cdot 40~\textrm{MHz}}{2 f_{\textrm{EL CLK}}}\right)$.

Therefore, for E-Link clock frequency $f_{\textrm{EL CLK}}=40~\textrm{MHz}$, the GBTxEMU may support up to 56 E-Links,
while for $f_{\textrm{EL CLK}}=80~\textrm{MHz}$ only 28.

The original GBTX ASIC offers a possibility to control the phase of the generated E-Link clock and the delay of the input E-Link data.
That enables adaptation to different values of the skew and length of the cable used for the FEE ASIC connection.
In the GBTxEMU, the E-Link clock phase may be adjusted with a resolution of 78.125~ps using the MMCM block to produce the output clock. 
The input data from the E-Link are delivered through the IDELAYE2 blocks  with run-time controllable 
delay (up to 2.496~ns with 78 ps resolution and with selectable clock edge)~\cite{cbm-prog-rep-2019-WUT}.
When used together, those functions enable reliable operation of the E-Links.

\section{Current use and results}
\begin{figure}[htbp]
	\centering %
	\includegraphics[height=2.1 cm]{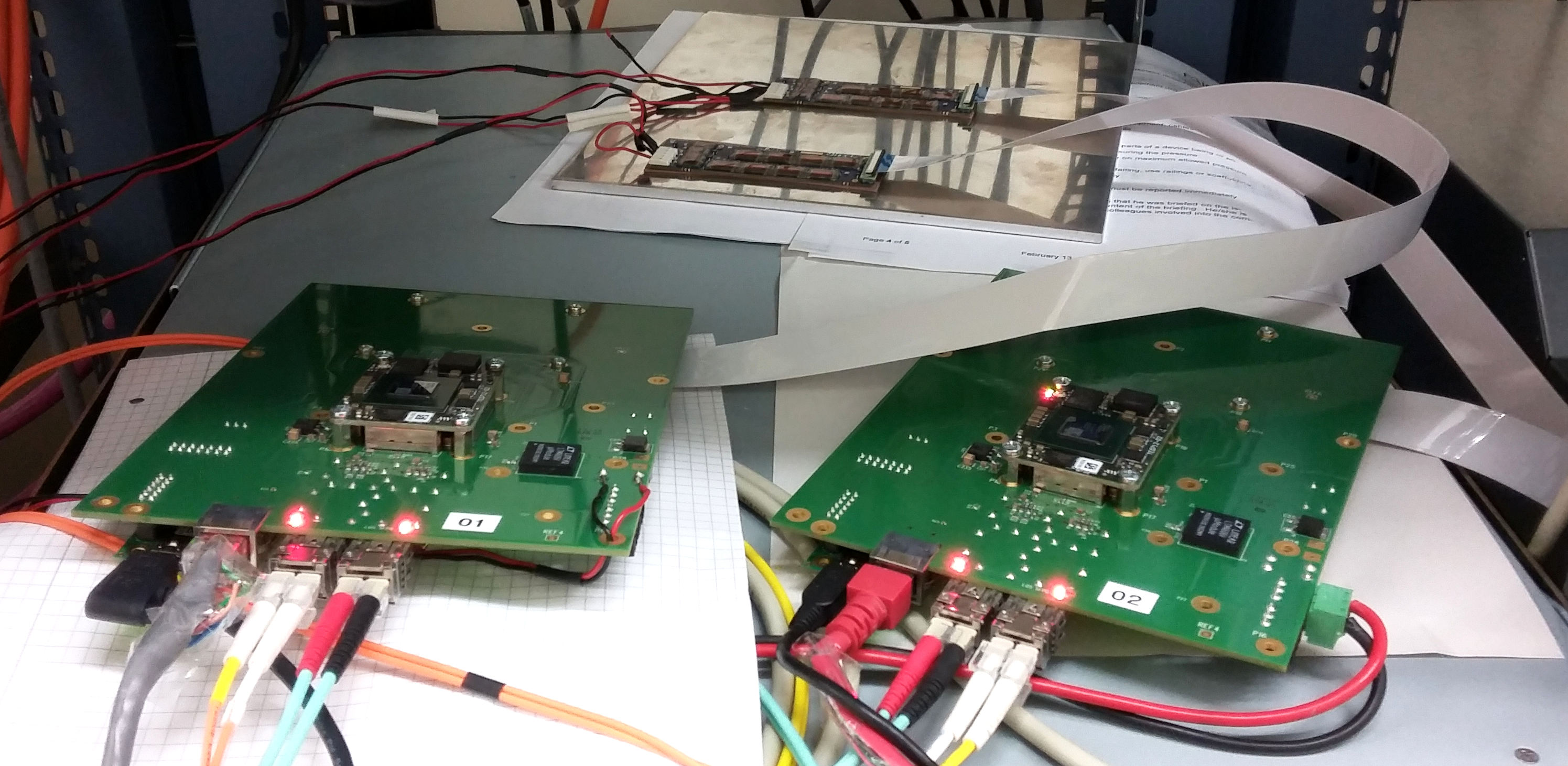}
	~
	\includegraphics[height=2.1 cm]{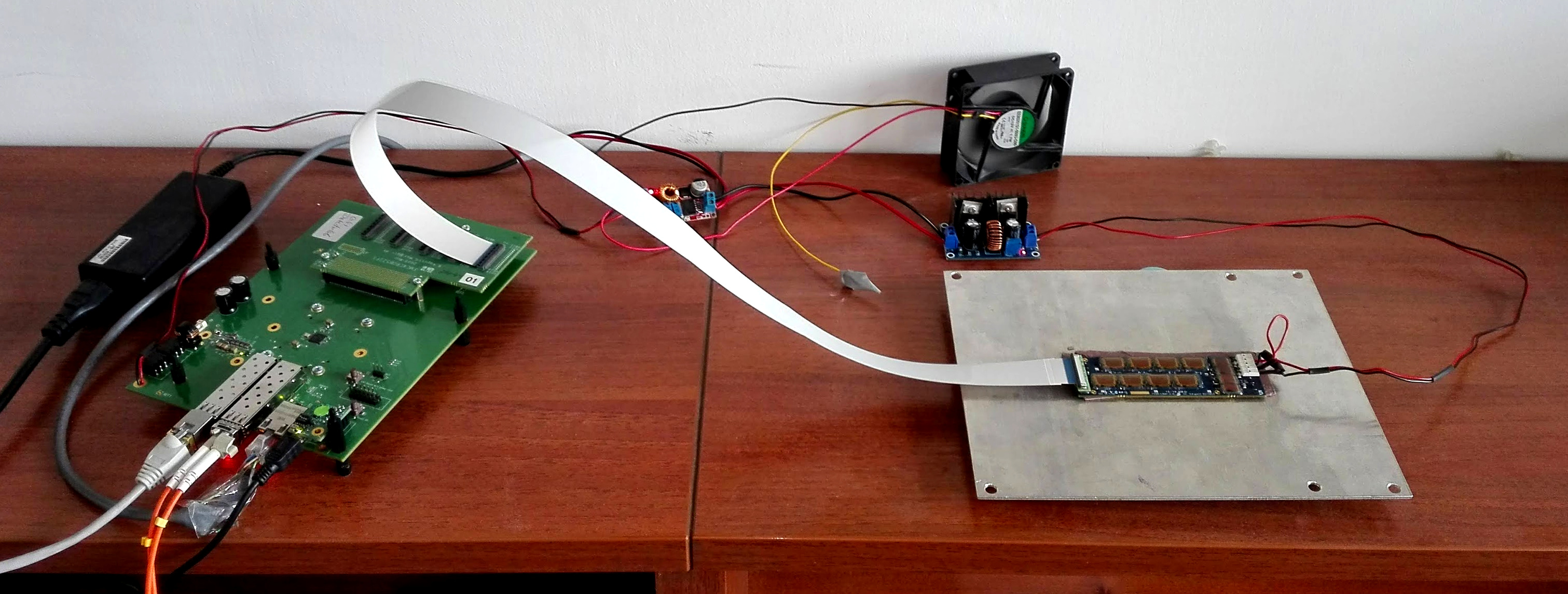}
	~
	\includegraphics[height=2.1 cm]{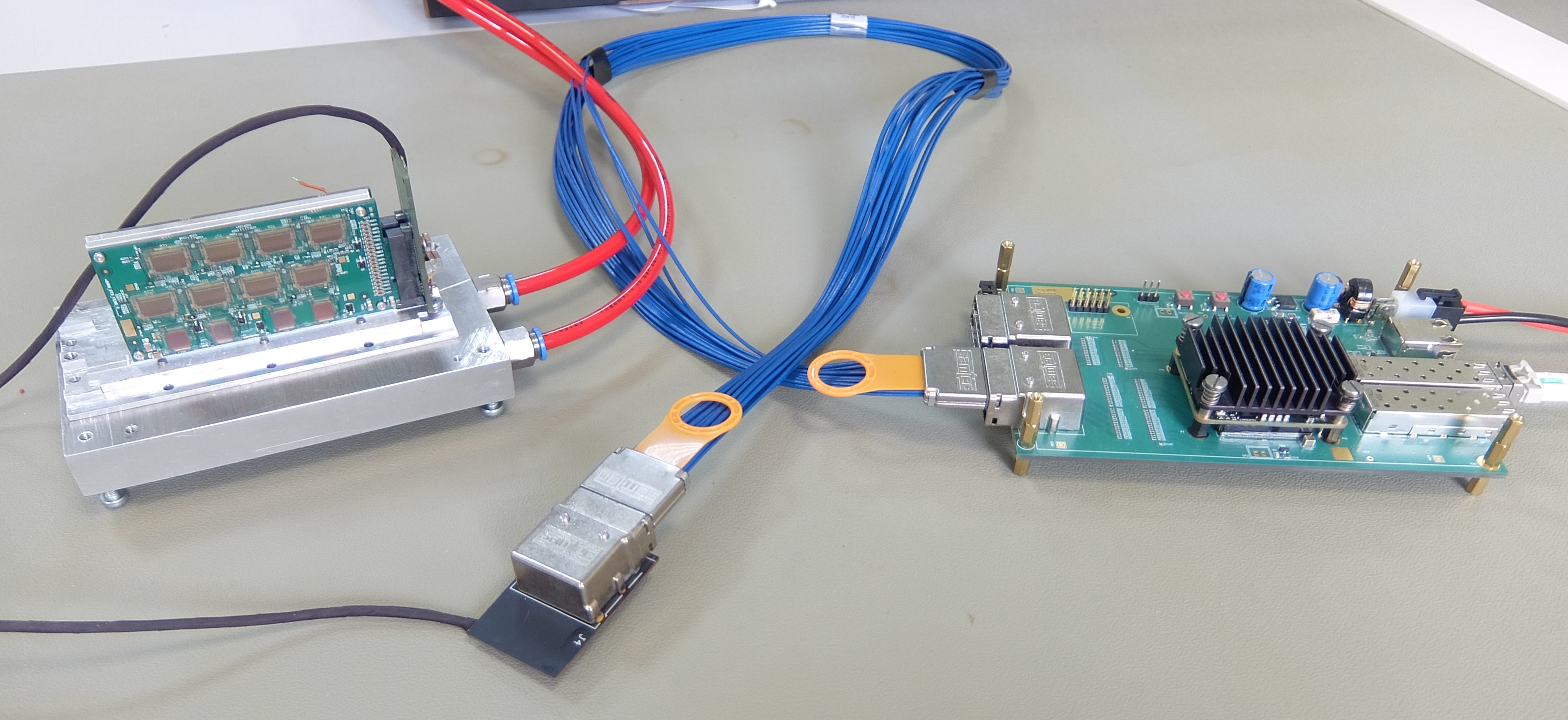}
	\caption{\label{fig:test-setups} Three test setups equipped with GBTxEMU boards.
	On the left -- the setup in GSI with 2 GBTxEMU v1 and two connected FEB-8 boards.
    In the center -- the setup at WUT with a single GBTxEMU v1 and connected FEB-8 board.
    On the right -- setup at JINR with a GBTxEMU v2 and connected BM@N STS FEB board.
}
\end{figure}

Different versions of the GBTxEMU have been used in test setups in WUT, GSI, and JINR (see Figure~\ref{fig:test-setups}).
The WUT setup is used mainly for development and testing purposes.
In GSI, the GBTxEMU has been successfully used for debugging the port of the GBT-FPGA developed for the CRI readout board.
The GBTxEMU provided extended diagnostics at the slave end of the GBT link.
In JINR, the GBTxEMU is used to develop the readout chain for the BM@N. 
It is also planned to be used in its final version~\cite{dementev_fast_2021}. Of course, the FPGA-based
design can not assure radiation hardness.
However, in situations where the highest throughput of E-Links is not needed, the board may be placed at some distance
from the on-detector electronics, outside of a high magnetic ﬁeld and radiation environment.
The 10-meter length copper cable connection between the front-end
electronics and GBTxEMU was proven to be stable in the BM@N STS project.

\section{Conclusions}

The GBTX emulator (GBTxEMU) appeared to be a useful tool enabling the
development of GBT-based readout systems in locations where the original
GBTX ASIC cannot be used. It has also shown its potential at debugging the GBT
links during porting of the GBT-FPGA IP core to the new FPGAs.
The GBTxEMU may be used in special versions of the GBT-based readout
chain, where radiation hardness is not essential. It may also be a basis for creating the GBT-controlled 
test setups for testing the FEE ASICs and modules connected via
GBTX-compatible E-Links.
The FPGA-based design is highly flexible, providing the end-users with the possibility 
to implement their own extensions.

\acknowledgments

The work has been partially supported by GSI, and partially by the statutory funds of Institute of Electronic Systems. This project has also received funding from the European Union’s Horizon 2020 research and innovation programme under grant agreement No 871072.

\bibliography{gbtxemu}
\bibliographystyle{JHEP}

\end{document}